\documentclass[twocolumn,superscriptaddress,amsmath,amssymb,aps,prb]{revtex4-2}
\usepackage{graphicx}% Include figure files
\usepackage{dcolumn}% Align table columns on decimal point
\usepackage{bm}% bold math
\usepackage{fixmath}% italic symbol
\usepackage[normalem]{ulem}
\usepackage[usenames,dvipsnames]{color}
\usepackage{wasysym}
\usepackage[colorlinks,bookmarks=true,citecolor=blue,linkcolor=blue,urlcolor=blue]{hyperref}

\begin{document}

\preprint{APS/123-QED}

\title{Real space representation of topological system: twisted bilayer graphene as an example}

\author{Jiawei Zang}
\email{jz3122@columbia.edu}
\affiliation{Department of Physics, Columbia University, 538 W 120th Street, New York, New York 10027, USA}
\author{Jie Wang}
\affiliation{Center for Computational Quantum Physics, Flatiron Institute, 162 5th Avenue, New York, NY 10010, USA}
\author{Antoine Georges}
\affiliation{Center for Computational Quantum Physics, Flatiron Institute, 162 5th Avenue, New York, NY 10010, USA}
\affiliation{Coll\`ege de France, 11 place Marcelin Berthelot, 75005 Paris, France}
\affiliation{CPHT, CNRS, \'Ecole Polytechnique, Institut Polytechnique de Paris, Route de Saclay, 91128 Palaiseau, France}
\affiliation{DQMP, Universit\'e de Gen\`eve, 24 Quai Ernest Ansermet, CH-1211 Gen\`eve, Switzerland}
\author{Jennifer Cano}
\affiliation{Center for Computational Quantum Physics, Flatiron Institute, 162 5th Avenue, New York, NY 10010, USA}
\affiliation{Department of Physics and Astronomy, Stony Brook University, Stony Brook, New York 11794, USA}
\author{Andrew J. Millis}
\affiliation{Department of Physics, Columbia University, 538 W 120th Street, New York, New York 10027, USA}
\affiliation{Center for Computational Quantum Physics, Flatiron Institute, 162 5th Avenue, New York, NY 10010, USA}

\begin{abstract}
We construct a Wannier basis for twisted bilayer graphene that is projected only from the Bloch functions of the twisted bilayer flat bands. The $C_3$ and $C_{2} \mathcal{T}$ symmetries act locally on the Wannier functions while the Wannier function charge density is strongly peaked at the triangular sites and becomes fully sublattice-polarized in the chiral limit. The Wannier functions  have a power-law tail, due to the topological obstruction, but most of the charge density is concentrated within one unit cell so that the on-site local Coulomb interaction is much larger than the further neighbor interactions and in general the Hamiltonian parameters may be accurately estimated from a modest number of Wannier functions. One exception is the  momentum space components of the single-particle Hamiltonian, where because of the topological obstruction convergence is non-uniform across the Brillouin zone. We observe, however, that mixed position and momentum space representations may be used to avoid this difficulty in the context of quantum embedding methods. Our work provides a new route to study systems with topological obstructions and paves the way for the future investigation of correlated states in twisted bilayer graphene, including studies of non-integer fillings and temperature dependence.
\end{abstract}
\maketitle

\emph{Introduction---}
The interplay of correlation physics and band topology is of great current interest~\cite{Tokura22} and calls for further development of theoretical methodologies. One important issue is the choice of basis. If a theory is solved exactly, the physical content is independent of the basis choice. However quantum many-body systems cannot in general be solved exactly and both the quality and the physical content of any approximate solution are affected by the choice of basis. In studying the physics of strongly interacting electrons in periodic potentials, spatially discretized representations such as Kanamori-Hubbard type models~\cite{HubbardNature13} have proven very useful because correlation physics typically involves strong local quantum and thermal fluctuations leading to local moment formation~\cite{Brandow77}, Mott transitions~\cite{RevModPhysMott}, orbital fluctuations, low temperature entropy release~\cite{Pomeranchuk1,Pomeranchuk2,Fwerner}, and related phenomena.  

Localized representations may be built from the Wannier functions derived~\cite{Marzari12} from Bloch functions provided by a band theory calculation. However, for topologically nontrivial bands the standard Wannierization procedure encounters difficulties~\cite{Wannier_obstructions1,Wannier_obstructions2,Soluyanov_2011,PhysRevB.102.075142}. In these situations, Wannier functions that transform locally under the relevant point symmetries are power-law rather than exponentially localized and while integrals such as those needed for determining Hamiltonian parameters exist, the integral for the mean square position uncertainty $\langle R^2\rangle$ is divergent so the standard Maximally Localized Wannier construction~\cite{Marzari12} cannot be directly applied.

Perhaps more importantly, in any Wannier description, the one-electron properties are described by a tight binding model, and one must generically include an infinite number of hopping parameters to exactly reproduce the band structure. For the usual exponentially localized Wannier functions the one electron Hamiltonian converges to the exact dispersion exponentially rapidly as more hoppings are included, and the convergence is uniform in momentum space. In the topologically obstructed case, the convergence is power-law in the number of included orbitals and, more seriously, the convergence of the single particle (hopping) parameters of the Hamiltonian is non-uniform in momentum space as will be described in detail below. The non-uniform convergence is related to constraints on the implementation of point symmetries in momentum and position space representations~\cite{Vafek_2021}. These difficulties have impeded the use of Wannier representations in studying correlation physics in topological bands. 

In this paper we use the example of twisted bilayer graphene (TBG)~\cite{tbg1,tbg2,Santos07,Bistritzer_2011} to show that even for topological bands useful Wannier representations may be constructed in which the Wannier functions faithfully represent important properties of the system and point symmetries act locally. Further, we observe that the real-space representation of the single-particle part of the Hamiltonian is not needed for many practical many-body calculations. As we demonstrate explicitly using the example of dynamical mean field theory, a mixed position/ momentum space representation can be employed, in which the kinetic energy is expressed in the momentum space basis of non-interacting eigenstates, so that all the topological features are exact and well preserved, while the interaction part may be expressed in position space and inherit convenient locality and symmetry properties from the Wannier functions~\cite{Georges96,Chen22,Beck_2022}. 

Specifically, in this paper we explicitly construct a Wannier basis for twisted bilayer graphene involving two triangular site-centered Wannier functions per unit cell derived from the two flatbands per spin per valley and show that these provide a physically intuitive and mathematically convenient real space picture. The two crucial point symmetries $C_2\mathcal{T}$ and $C_3$ act locally on the Wannier functions we construct, and within a unit cell their charge density profile corresponds closely to that found in scanning probe experiments~\cite{STM_Abhay}. Although the Wannier functions have a power-law decay arising from the topological obstruction, we find that they are in practice very localized, leading to a computationally convenient representation of the interactions in which the  on-site terms are much bigger than the first or farther neighbor terms.  

The rest of this paper is organized as follows. We first summarize relevant aspects of the physics of twisted bilayer graphene and then review the continuum model that is believed to describe the low energy physics and summarize the symmetry of the eigenstates. Next we construct the Wannier functions and present a detailed analysis of their symmetry properties and spatial structure. Then we calculate the interaction and show how to use a mixed real and momentum space representation to study correlated states. Finally we present a summary and broader outlook.

\begin{figure}[htbp]
	\centering
	\includegraphics[width=1\linewidth]{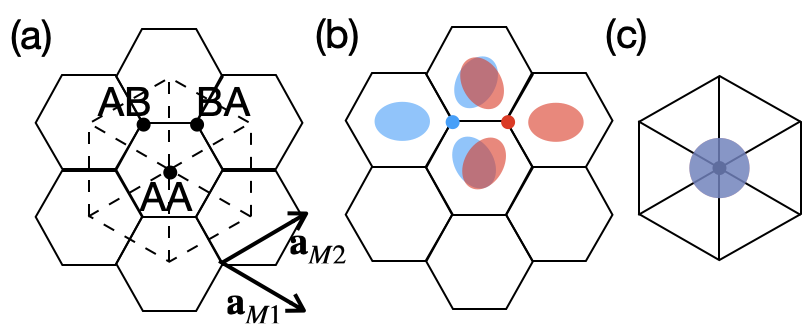}
	\caption{(a) Representation of hexagonal moir\'e lattice with triangular (hexagon-centered) AA sites and hexagonal (AB/BA) sites indicated. Solid lines show the hexagonal lattice with vertices at the AB/BA sites and dashed lines show the triangular lattice with vertices at the AA sites. (b) Sketch of AB/BA site-centered Wannier functions. Blue and orange dots show the hexagonal AB and BA sites respectively; shaded blue and orange ellipses show the charge densities of the corresponding two Wannier functions and demonstrate the  spatially extended ``fidget spinner'' shape required for the charge density of the Wannier functions to correspond to the physical charge density. (c) Sketch of the two Wannier functions centered at the triangular AA site. At the coarse grained level shown here the charge densities of the two orbitals are the same.}
	\label{fig:real_space}
\end{figure}

\emph{Twisted bilayer graphene---}
Twisted bilayer graphene (TBG) is a system comprised of two graphene sheets stacked one on top of the other at a small twist angle. At commensurate twist angles one finds a moir\'e pattern characterized by a hexagonal unit cell as shown in panel (a) of Fig.~\ref{fig:real_space}. The hexagonal unit cell may be very large compared to the lattice constant of the underlying graphene. At the experimentally interesting carrier concentrations and appropriate small twist angles, calculations~\cite{Santos07,Bistritzer_2011} indicate that there are eight relevant bands, two per spin per valley. The absence of spin-orbit coupling and the exponential suppression of intervalley mixing in large unit cells  means that the spin and valley quantum numbers may be thought of as internal quantum numbers attached to two bands of electrons that may be described as eigenstates of  a `continuum' model \cite{Bistritzer_2011} involving a Dirac dispersion in each plane and a spatial periodicity defined by the interplane coupling.  These two  bands are often referred to as `flatbands' because for appropriate twist angles they are well separated from the other bands and their dispersion can be very weak relative to the dispersion of other bands. The flatbands of TBG are of intense current interest for the wide variety of novel phases and correlated electron physics they host \cite{tbg1,tbg2}.

Density functional theory \cite{Santos07} and scanning probe experiments \cite{STM_Abhay} confirm that for almost all momenta in the moir\'e Brillouin zone the charge density of the flat band electrons is concentrated at the triangular (hexagon-center) AA sites, suggesting that the  real-space description should be based on two Wannier orbitals (each with spin and valley quantum numbers) that are centered at the triangular sites, transform appropriately under the point symmetries and are constructed from the flat band states.

However, it is known that the flat bands in TBG exhibit a topological obstruction that prohibits the construction of exponentially-localized Wannier functions on which the relevant symmetries act locally~\cite{Wannier_obstructions1,Wannier_obstructions2}. Several strategies have been proposed to resolve the problem. One of them is to avoid the problem and work entirely in momentum space~\cite{Ground_State_integer}. This approach is suitable for constructing ground states and for Hartree-Fock based studies, but is not convenient for several many-body methods beyond Hartree-Fock. Building Wannier functions from a large number of bands~\cite{morebands,tbg_heavy_fermion,XiDai} removes the topological obstruction at the cost of greatly widening the energy range that must be considered and complicating the theoretical model.

The topological obstruction may also be avoided by building exponentially-localized Wannier functions centered at the hexagonal AB and BA sites \cite{hexagonal_LiangFu,Hexagonal_Vafek, Vafek_2021}, as shown in Fig.~\ref{fig:real_space}(b). The $C_2\mathcal{T}$ symmetry of TBG acts non-locally on these states. Additionally, the need to capture the triangular site centered charge densities using hexagon-center Wannier functions leads to a spatially extended  ``fidget spinner'' shape \cite{hexagonal_LiangFu,Hexagonal_Vafek} that leads to highly non-local real space interactions~\cite{oskar_prl19}. Alternative representations of the local interaction physics are therefore desirable.

\emph{Continuum model and its symmetries.---}
The continuum Hamiltonian $H^{\eta}({\bm r})$ for one valley $\eta=\pm$ is a $4 \times 4$ matrix with the basis (A1, B1, A2, B2), corresponding to the two layers 1, 2 and two sublattices A, B of monolayer graphene. Here we follow the convention of Ref.~\cite{hexagonal_LiangFu}, in which $H^{\eta}({\bm r})$ is written as 
\begin{equation}
H^{\eta}({\bm r})=\left(\begin{array}{cc}h_{1,-\theta / 2}^{\eta}({\bm r}) & T^{\eta}({\bm r}) \\ T^{\eta\dagger}({\bm r}) & h_{2, \theta / 2}^{\eta}({\bm r})\end{array}\right),
\label{eq:Hamiltonian}
\end{equation}
where the motion of an electron within one plane is described by 
\begin{equation}
h_{l, \mp \theta / 2}^{\eta}=-\hbar v\left[R\left(\pm \frac{\theta}{2}\right)\left(-i \nabla-{\bm k}_{\eta}^{l}\right)\right] \cdot\left(\eta \sigma_{x}, \sigma_{y}\right),
\label{eq:h}
\end{equation}

Here, the $\mathbf K^{1,2}_{\eta} = R_{\mp\frac{\theta}{2}}\mathbf K_{\eta}$ are the Dirac points of layer $1,2$, obtained by rotating the monolayer graphene Dirac point $\mathbf K_{\eta}$ by angle $\mp\theta/2$, respectively. In our numerical calculations we fix the velocity $v$ and a single-plane lattice constant $a$ as $-\hbar v/a=2135.4$ meV and $a=0.246$ nm and set the twist angle to $\theta=1.05^\circ$ \cite{hexagonal_LiangFu}.

The interlayer tunnelling $T$ is given by
\begin{equation}
		T^\eta({\bm r})=\sum_{j=1}^{3}\left(\begin{array}{cc}u_{0} & u_{1} e^{-i\eta(j-1) \frac{2\pi}{3}} \\ u_{1} e^{i\eta(j-1)\frac{2\pi}{3}} & u_{0}\end{array}\right) e^{-i \eta \mathbf{g}_{j} \cdot {\bm r}},
		\label{eq:T}
\end{equation}
where $\mathbf g_{j=1,2,3}$ are the three smallest reciprocal lattice vectors of TBG. $T$ involves two parameters: $u_0$ describing same-sublattice interlayer tunnelling and $u_1$ describing inter-sublattice interlayer tunnelling. For most of the numerical calculations presented in this paper we choose $u_0= 79.7$ meV, $u_1= 97.5$ meV. In the {\it chiral limit}, where $u_0=0$, the model has special properties~\cite{chiral,Wang21,WangPRL21,WangPRL22,EslamPRL22,Geardo21}. To study the chiral limit we set $u_0=0$ keeping the other parameters unchanged.

The eigenstates of Eq.~(\ref{eq:Hamiltonian}) are four-component spinors $\psi_{m{\bm k}}({\bm r})$ labelled by a band index $m$ and a momentum ${\bm k}$ in the first moir\'e Brillouin zone. Throughout this work, we use $\bm l, \bm\sigma, \bm v$ to represent the Pauli matrices acting on layer, sublattice, and valley indices respectively. The spinor components $\psi_{m{\bm k}}^{\eta X}({\bm r})$ of an eigenstate ($X=(A1,B1,A2,B2)$) may be expanded as:
\begin{equation}
    \psi_{m{\bm k}}^{\eta X}({\bm r})=\sum_G \frac{1}{\sqrt{V}}b^{\eta  X}_{m\mathbf{G}}({\bm k})e^{i({\bm k}+\mathbf{G})\cdot {\bm r}}
\end{equation}
where $\bm G$ is the moir\'e reciprocal lattice vector. It is important to note that we have made a gauge choice such that $\bm k=0$ corresponds to the moir\'e $\Gamma$ point. 

The system has $C_2\mathcal{T}$ and $C_3$ symmetry and admits an extra chiral symmetry $C$ in the chiral limit $u_0=0$~\cite{chiral,Wang21}. The explicit representation for $C_2T$ is $\sigma_{x} \hat{R}_{z}(\pi)\mathcal{K}$, a combination of sublattice swap, $\bm r\rightarrow-\bm r$ and time-reversal (complex conjugation). The representation of $C_3$ is $W^\dagger({\bm r}) e^{i\frac{2\pi}{3}\sigma_z v_z} \hat{R}_{z}(2 \pi / 3)W({\bm r})$, where $W({\bm r})=e^{-i {\bm k}_{\eta}^{l} \cdot {\bm r}}$
moves the rotation center of each layer to its Dirac point. The chiral symmetry, present only at the chiral limit, is simply $C=\sigma_z$.

Thus, eigenstates of the symmetries above satisfy the following constraints:
\begin{eqnarray}
    C_2\mathcal{T}\psi_{m{\bm k}}^{\eta}({\bm r}) &=& \sigma_{x} \psi_{m{\bm k}}^{\eta *}(-{\bm r}) = e^{i \alpha^\eta_{m{\bm k}}} \psi_{m{\bm k}}^{\eta}({\bm r});
    \label{eq:bloch_symmetryC2T}\\
    C_3\psi_{m{\bm k}}^{\eta}({\bm r}) &=& e^{i\beta^\eta_{m{\bm k}}}\psi^{\eta}_{mC_3{\bm k}}({\bm r})\label{eq:bloch_symmetry}
\end{eqnarray}
In the chiral limit, the chiral eigenstate in addition satisfies,
\begin{equation}
    \begin{aligned}
    &C\psi_{m {\bm k}}^{\eta}({\bm r})=\sigma_z \psi_{m {\bm k}}^{\eta}({\bm r})=e^{i 
    \zeta^\eta_{{\bm k}}} \psi_{\bar{m},{\bm k}}^{\eta}({\bm r}),
    \end{aligned}
    \label{eq:C}
\end{equation}
where $\bar 1 = 2$ and $\bar 2 = 1$: the chiral symmetry flips the band index because by definition $\{H^{\eta}({\bm r}),C\}=0$.

%  I numerically find that 
%  \begin{equation}
%     \begin{aligned}
%     &\psi_{n k}^{A2\eta}(r)=\psi_{n k}^{A1\eta}(r)e^{-i\pi/2}\\
%     &\psi_{n k}^{B2\eta}(r)=\psi_{n k}^{B1\eta}(r)e^{i\pi/2}\\
%     \end{aligned}
% \end{equation}

\emph{Construction of the Wannier function and symmetry---}
Wannier functions are generalized Fourier transforms of Bloch states specified by a ${\bm k}$ dependent unitary matrix $M$ whose dimension is the number of Bloch bands used to make up the Wannier function. We focus here on the two four-component Wannier spinors made from Bloch states in the two flatbands:
\begin{equation}
	\begin{aligned}
	&w^\eta_{n{\bm R}}({\bm r})=\frac{1}{\sqrt{N}} \sum_{{\bm k}} e^{-i {\bm k} \cdot {\bm R}}\sum_{m=1,2} \psi^\eta_{m {\bm k}}({\bm r}) M^\eta_{m n}({\bm k})\\
	&M^\eta ({\bm k})=\left(\begin{array}{cc}a^\eta_{k} & e^{i \phi^\eta_{k}} b_{k}^{\eta*} \\ b^\eta_{k} & -e^{i \phi^\eta_{k}} a_{k}^{\eta *}\end{array}\right),~|a_{k}^{\eta }|^{2}+|b^\eta_{k}|^{2}=1
	\label{eq:wannier}
\end{aligned}
\end{equation}
where ${\bm R}=m_1\mathbf{a}_{M1}+m_2\mathbf{a}_{M2}$ ($m_1,m_2\in Z$) is the moir\'e lattice site (Wannier center), with $\mathbf{a}_{M1,M2}$ the primitive vectors shown in Fig.~\ref{fig:real_space}(a). The localization and symmetry properties are tuned by the matrix $M^\eta({\bm k})$ and for simplicity we assume that the unitary matrix acts as the unit matrix in spinor space: in other words each of the four spinor components is Wannierized in the same way.
%One can take the advantage of the phase freedom of Bloch states to optimize the Wannier function, by choosing a proper $U^\eta({\bm k})$ that minimizes its spatial spreading.

To make the Wannier construction well defined, the phases of the Bloch functions $\psi_{m{\bm k}}({\bm r})$ must be specified. Since the density is strongly peaked at the triangular sites \cite{STM_Abhay}, we choose the convention that for each ${\bm k}$ point the B1 component of each spinor Bloch state, $\psi^{\eta B1}_{m=1,2;{\bm k}} ({\bm r})$ is real and positive at ${\bm r}=0$, {\it i.e.} at the AA site. This condition, which may be thought of as a choice of gauge, fixes the symmetry phases $\alpha_{m {\bm k}}^{\eta},\beta_{m{\bm k}}^{\eta}$ and $\zeta_{{\bm k}}^{\eta}$ defined in Eqs.~(\ref{eq:bloch_symmetryC2T}), (\ref{eq:bloch_symmetry}) and (\ref{eq:C}).

\begin{figure}[t]
	\centering
	\includegraphics[width=1.0\linewidth]{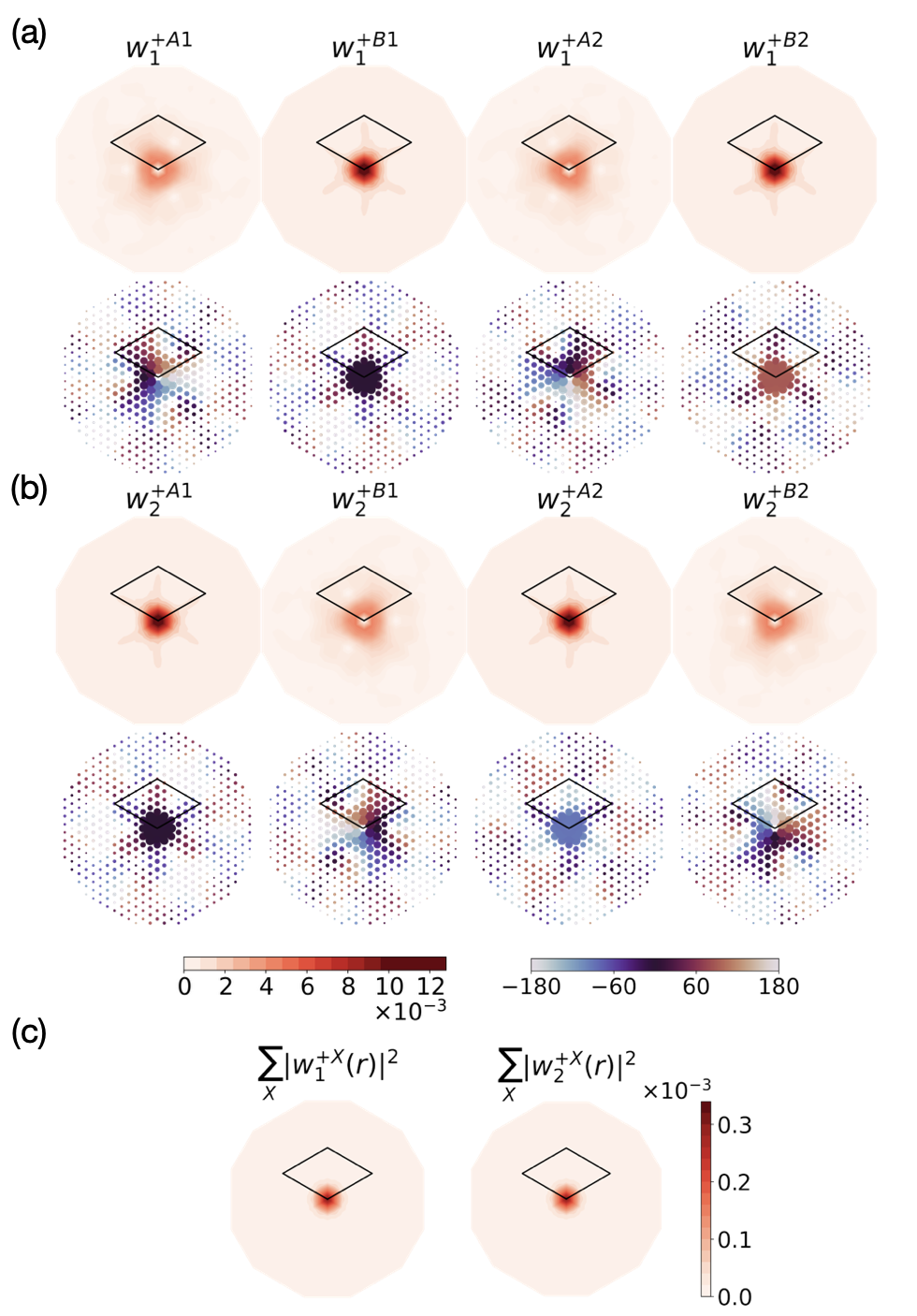}
	\caption{(a, b) Two Wannier functions $w_{1 {\bm R}}^{+}({\bm r})$ and $w_{2 {\bm R}}^{+}({\bm r})$ for valley $+$ at ${\bm R}=0$. The first line shows the absolute amplitude of the four components of $w_{1 {\bm R}}^{+}({\bm r})$ and the second line shows the corresponding phase of the envelope function defined as $W({\bm r})w_{1 {\bm R}}^{+}({\bm r})$. The third and fourth lines are for $w_{2 {\bm R}}^{+}({\bm r})$. The Wannier function is centered at the triangular lattice site. (c) Position dependence of the charge densities $\rho_{m=1,2}({\bm r})=w_{m}^{+\dagger}({\bm r})w_{m}^{+}({\bm r})$ of the two  Wannier functions in the $+$ valley. In all plots the moir\'e unit cell boundary is shown as a black rhombus and the AA site is at the center of the circle.}
	\label{fig:wannier_triangular}
\end{figure}

We then choose $M^\eta({\bm k})$ such that the resulting Wannier functions transform as follows under $C_3$ and $C_{2} \mathcal{T}$ symmetry operations,
\begin{equation}
	\begin{aligned}
	    &C_3 w_{n{\bm R}}^{\eta}({\bm r}) =e^{-i\frac{2\pi}{3}n\eta}w_{nC_3{\bm R}}^{\eta}({\bm r})\\
	   &C_{2} \mathcal{T} w^\eta_{1{\bm R}}({\bm r})=\sigma_{x}w^{\eta *}_{1{\bm R}}(-{\bm r})=w^\eta_{2,-{\bm R}}({\bm r})
	\end{aligned}
\end{equation}

Specifics of the Wannier construction are given in Appendix~\ref{sec:AppendixC}. In brief, the symmetry requirements and the gauge choice that fixes the overall phase of the wave functions at each $k$ fixes the phase $\phi_k$ and the ratio of the $a_k$ and $b_k$ in the unitary operator $M$ in Eq.~(\ref{eq:wannier}). One might further adjust the $k$ dependence of the magnitudes of $a_k,b_k$ to localize the Wannier function. However, the meaning of localization requires discussion. As will be seen, the Wannier functions decay as $1/r^2$ so the integral in the conventional localization criterion spread $\int d^2r~r^2~w(\mathbf r)^2$ is divergent. We suggest that one may use instead as a simple localization criterion the magnitude $f$ of the Wannier function weight in the primitive cell: $f({\bm r})=\int_{\hexagon} |w_{ {\bm R}=0}({\bm r})|^2$. The $a_k$ can then be adjusted to maximize $f({\bm r})$. We  find however that the simple choice of a $k$-independent $a$ given in Appendix ~\ref{sec:AppendixC} already leads to a quite localized function with $f({\bm r})\approx 84\%$, and we have therefore not attempted a further optimization. 

Interestingly, at the chiral limit, the Wannier functions simplify to
\begin{equation}
	\begin{aligned}
	w^\eta_{1{\bm R}}({\bm r})&=\frac{1}{\sqrt{N}} \sum_{{\bm k}} e^{-i {\bm k} \cdot {\bm R}} \frac{\psi^\eta_{1 {\bm k}}({\bm r})+\psi^\eta_{2 {\bm k}}({\bm r})}{\sqrt{2}}\\
	&=\frac{1}{\sqrt{N}} \sum_{{\bm k}} e^{-i {\bm k} \cdot {\bm R}} \frac{1-\sigma_z}{\sqrt{2}}\psi^\eta_{1 {\bm k}}({\bm r})\\
	w^\eta_{2{\bm R}}({\bm r})&=\frac{1}{\sqrt{N}} \sum_{{\bm k}} e^{-i {\bm k} \cdot {\bm R}} e^{i\alpha^\eta_{1{\bm k}}}\frac{\psi^\eta_{1 {\bm k}}({\bm r})-\psi^\eta_{2 {\bm k}}({\bm r})}{\sqrt{2}}\\
	&=\frac{1}{\sqrt{N}} \sum_{{\bm k}} e^{-i {\bm k} \cdot {\bm R}} \frac{1+\sigma_z}{\sqrt{2}}e^{i\alpha^\eta_{1{\bm k}}}\psi^\eta_{1 {\bm k}}({\bm r})\\
\end{aligned}
\end{equation}
The two Wannier functions become fully sublattice-polarized, so that  $w_{1{\bm R}}^{\eta}({\bm r})$ and $w_{2{\bm R} }^{\eta}({\bm r})$ only have B and A sublattice components, respectively, and have the following chiral symmetry:
\begin{equation}
	\begin{aligned}
	Cw^\eta_{1{\bm R}}({\bm r})&=-w^\eta_{1{\bm R}}({\bm r}),~Cw^\eta_{2{\bm R}}({\bm r})&=w^\eta_{2{\bm R}}({\bm r})
\end{aligned}
\end{equation}

The amplitudes and phases of the four constructed Wannier functions are shown in Fig.~\ref{fig:wannier_triangular} away from the chiral limit; analogous plots for the chiral case are shown in Appendix~\ref{sec:AppendixC}. 

\emph{Wannier function properties---}
The topological obstruction in TBG prohibits the construction of exponentially-localized Wannier functions on which the relevant symmetries act locally \cite{Wannier_obstructions1,Wannier_obstructions2}.

\begin{figure}[b]
	\centering
	\includegraphics[width=1.0\linewidth]{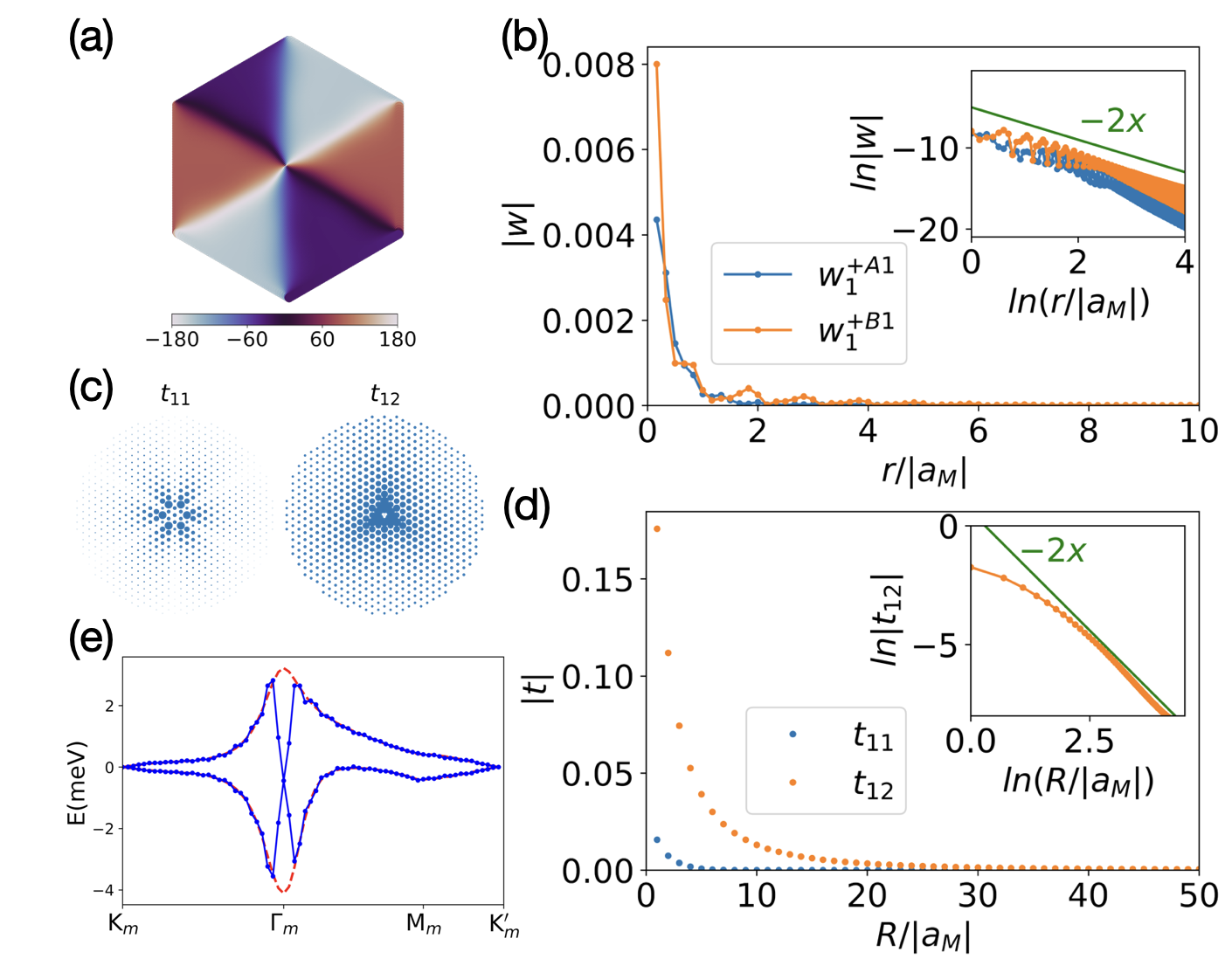}
	\caption{(a) Phase of the A1 component of $\psi_{1 {\bm k}}^{\eta}({\bm r}=0)$ plotted in the moir\'e Brillouin zone for valley $\eta=+$. (b) $|w_{1,R=0}^{+A 1}(r)|$ and $|w_{1,R=0}^{+B 1}(r)|$ along the y direction. Inset: log-log plot of $|w_{1,R=0}(r)|$.  (c) $|t_{11}^{0R}|$ and $|t_{12}^{0R}|$ in real space. Each point represents a lattice site $R$, and the size of the points represents the size of $|t|$. (d) $|t_{11}^{0R}|$ and $|t_{12}^{0R}|$ along the y direction. Inset: log-log plot of $|t_{12}^{0R}|$. The results are for valley $\eta=+$. The valley $\eta=-$ case can be obtained by time reversal symmetry. (e) The band structure. The blue bands show the tight binding parameters with truncation range $|R|=15 a_M$, while the red bands are calculated from the continuum model.}
	\label{fig:phase_and_decay}
\end{figure}

Consistent with those arguments, the Wannier functions we construct have a power-law decay, with amplitude falling as $1/r^2$ for large distances from the Wannier center. Mathematically, the decay arises from an incompatibility between the phase fixing condition that the $B1$ component of wave function at $\mathbf r=0$ is real and the requirement following from the $C_3$ and $C_{2} \mathcal{T}$ symmetries that all four components of the  $\Gamma$ point Bloch function vanish at ${\bm r}=0$, i.e. $\psi^{\eta}_{m\Gamma}(0)=0$.  This combination causes the phase of the wave function to wind by $2\pi$ around the $\Gamma$ point as shown in Fig.~\ref{fig:phase_and_decay}(a), implying a vortex-like singularity which is not removed by the unitary matrices specifying the Wannier function. This vortex-like singularity leads to the $r^{-2}$ power-law decay in the Wannier function shown in Fig.~\ref{fig:phase_and_decay}(b). The main part of this panel shows the amplitude of the $A1$, $B1$ components of the first Wannier function along the y direction, $\left|w_{1,{\bm R}=0}^{+}((0,y))\right|$ and the inset presents a log-log plot confirming the power-law decay. 
However as can be seen from the main part of Fig.~\ref{fig:phase_and_decay}(b) the Wannier function is still quite concentrated in one unit cell, with the amplitude  quickly decaying to be smaller than $2\%$ for $r>2|a_m|$. The weight in the primitive cell  $f({\bm r})=\int_{\hexagon} |w_{ {\bm R}=0}({\bm r})|^2$ is found to be $\approx 84\%$. We comment that the phase discontinuity always exists as it is guaranteed by the topological obstruction from the $C_3$ and $C_2T$ symmetry~\cite{Wannier_obstructions1,Wannier_obstructions2}; the particular gauge fixing choice determines where this discontinuity occurs.

As seen in panel (b) of Fig.~\ref{fig:wannier_triangular} the two Wannier functions on a given site have exactly the same charge density: $\rho^+_{n=1,2 ;{\bm R}}({\bm r})\equiv w_{n{\bm R}}^{+\dagger}({\bm r})w_{n{\bm R}}^{+}({\bm r})=\rho_{n{\bm R}}^-({\bm r})$ ($w^{\eta\dagger}_{n{\bm R}}({\bm r})w^\eta_{n{\bm R}}({\bm r})$ denotes the scalar product of the two spinors). The Wannier functions are orthogonal by construction: $\langle w_{n {\bm R}}({\bm r})|w_{ n^\prime\mathbf{R^\prime}}({\bm r})\rangle=\delta_{{\bm R},{\bm R}^\prime}\delta_{n,n^\prime}$ where the $\langle\rangle$ denotes the integral over ${\bm r}$ of the scalar product of the two spinors and the orthogonality arises from structure in the two spinor components, as is straightforwardly seen from the full sublattice polarization of the chiral limit wave functions.

\emph{Hamiltonian parameters---}
The projection of the energy bands onto the Wannier basis yields tight-binding hopping parameters $t^{\eta;{\bm R}^\prime,{\bm R}}_{n^\prime,n}$ ($\eta$ labels the valley and $n,n^\prime$ the orbital), which are calculated as follows,
\begin{equation}
    \begin{aligned}
    &t^{\eta {\bm R}^{\prime} {\bm R}}_{ n^{\prime} n }=\frac{1}{N} \sum_{{\bm k}} e^{i {\bm k} \cdot({\bm R}^{\prime}-{\bm R})}\left[M^{\eta \dagger}({\bm k}) \left(\begin{array}{cc}E^\eta _{1k} & 0 \\ 0 & E^\eta _{2k}\end{array}\right) M^\eta ({\bm k})\right]_{n^{\prime} n}\\
    &=\sum_{{\bm k}}\frac{e^{i {\bm k} \cdot({\bm R}^{\prime}-{\bm R})}}{2N}  \\ ~&\left(\begin{array}{cc}E^\eta_{1k}+E^\eta_{2k}& \sqrt{2}e^{i \phi^\eta_{k}}b_k^{\eta*}(E^\eta_{1k}-E^\eta_{2k}) \\ \sqrt{2}e^{-i \phi^\eta_{k}}b^\eta_k(E^\eta_{1k}-E^\eta_{2k}) & E^\eta_{1k}+E^\eta_{2k}\end{array}\right)_{n^{\prime} n}
    % &U^{k}=\left(\begin{array}{cc}1 & e^{i \phi_{k}} \\ 1 & -e^{i \phi_{k}}\end{array}\right) / \sqrt{2}\\
    \end{aligned}
\end{equation}
The hopping amplitude $t_{nn}$  between the same orbitals decays exponentially with distance because the component in the Fourier transformation doesn't have a discontinuity, while the hopping amplitude $t_{n\bar{n}}$ between different orbitals has a power-law decay due to the discontinuity of the phase in the Fourier transformation. The variation of  hopping amplitudes with relative position is  shown in panel (c) of Fig.~\ref{fig:phase_and_decay}. Panel (d) presents line cuts demonstrating the different spatial dependence of the hopping amplitudes and confirming the power-law decay of $t_{12}$.

The power-law decay of the hopping amplitudes means that the convergence of the k-space dispersion with number of hopping amplitudes is only power-law. More importantly, the convergence of the hopping amplitude is non-uniform in momentum space. Panel (e) of Fig.~\ref{fig:phase_and_decay} shows the band structure calculated from hopping amplitudes up to a range $|R|=15 a_{M}$, as shown in (c). With this level of truncation the bands at most momenta are very well reproduced, but the bands near the $\Gamma$ point are poorly reproduced. As the number of retained coefficients is increased the range of momenta around $\Gamma$ where the truncated tight binding model fails decreases, but at any finite truncation respecting lattice symmetry there is a two-fold degeneracy at the $\Gamma$ point, as shown in Fig.~\ref{fig:phase_and_decay}(e).  

The projection of the bare Coulomb interaction onto the Wannier basis gives the four-center Coulomb parameters $\hat{U}$ and $\hat{V}$. The required integrals involve four wave functions and a spatially decaying kernel and are absolutely convergent. The on site density-density interactions $U$ and exchange interactions $J$ are defined in terms of an atomic-scale interaction (we use the bare Coulomb interaction $V({\bm r}-{\bm r}')=\frac{e^{2}}{\epsilon\left|{\bm r}-{\bm r}'\right|}$ with screening parameterized by a dielectric constant $\epsilon$ ) as follows:
\begin{equation}
    \begin{aligned}
    \hat{U}_{n{\bm R},n'{\bm R}'}^{\nu\nu'}&=\frac{1}{2} U_{n{\bm R},n'{\bm R}'} \hat{n}^\nu_{n{\bm R}}  \hat{n}^{\nu'}_{n'{\bm R}' }\\
		U_{n{\bm R},n'{\bm R}'}&=%\sum_{XX'}
		\int d^{2} r d^{2} r' %|w_{n,i}^{\nu X}(r)|^2 
		w^{\nu\dagger}_{n{\bm R}}(r)w^{\nu}_{n{\bm R}}(r)V\left({\bm r}-{\bm r}'\right) 
			w^{\nu^\prime\dagger}_{n^\prime {\bm R}^\prime}(r)w^{\nu^\prime}_{n^\prime {\bm R}^\prime}(r)\\
    \hat{V}_{nn'}^{\nu\nu'}&=\frac{1}{2} J a_{n{\bm R} }^{\nu\dagger}  a
		_{n'{\bm R}}^\nu a_{n'R }^{\nu'\dagger} a_{n{\bm R}}^{ \nu'}\\
		J&=%\sum_{XX'}
		\int d^{2} r d^{2} r' w_{n{\bm R}}^{\nu \dagger}(r) w^{\nu }_{\bar{n}{\bm R}}(r)V({\bm r}-{\bm r}')
		w_{\bar{n}{\bm R}}^{\nu'\dagger}(r') w^{\nu'}_{n{\bm R}}(r') ,
    \end{aligned}
\end{equation}
Here $\hat{n}^\nu_{n{\bm R}}=a^{\nu\dagger}_{n{\bm R}}a^\nu_{n{\bm R}}$ is the orbital density operator in the Wannier basis, $\nu=\sigma,\eta$ includes spin and valleys, $\bar{n}$ is the orbital that is not $n$ and we recall that the $w^\dagger w$ notation indicates the scalar product the two four-component spinors. Note that the contributions of any atomic-scale on-site interaction ($U$ on a carbon $p_z$) are small by a factor of the order of the number of atoms in a moir\'e unit cell and are not considered here.

\begin{table}[ht]
\caption{\label{table}%
Interaction for the Wannier orbitals in units of $e^{2} /\left(\epsilon L_{M}\right)$.}
\begin{ruledtabular}
\begin{tabular}{cccc}
$R'$ & $0$ & $1$ & $2$ \\ \hline
${U}_{10,1R'}^{\nu\nu'}$ & 4.25 & 0.99 & 0.54  \\ 
${U}_{10,2R'}^{\nu\nu'}$ & 4.25 & 0.99 & 0.54 \\
\end{tabular}
\end{ruledtabular}
\end{table}

We find that the density-density interactions between the same orbitals and different orbitals are the same and the exchange interactions zero within numerical error. The onsite interaction is seen to be much larger than the further neighbor interactions. The identity of the different density-density interactions follows from the equality of the charge densities of the different orbitals while the vanishing of the exchange interactions comes from the interplay of the spinor wave functions and is most easily seen in the chiral limit as a consequence of the full sublattice polarization of the two wave functions. The density-density interaction parameters for onsite ($R'=0$), nearest ($R'=1$) and next nearest neighbors ($R'=2$) are shown in Table~\ref{table}.

\emph{Correlated states---} 
The onsite density-density interaction is much bigger than other interactions. Therefore as a first approximation, one may model TBLG as a multi-orbital Hubbard model using only this onsite interaction.
%\begin{equation}
%\begin{aligned}
% H_{int}&=\sum_{i}\frac{1}{2} U \sum_{n,\sigma} n^\sigma_{ni} \sum_{[p,\sigma']\neq[n,\sigma]}  n^{\sigma'}_{pi }-3.5Un^\sigma_{ni}\\
%   \hat{H}_{int} =\sum_{i}\frac{1}{2} U_0 \hat{N}_i(\hat{N}_i-8), ~\hat{N}_i= \sum_{n,\nu} n^\nu_{ni},
%\end{aligned}
%\label{eq:Hint}
%\end{equation}

Writing the model in a mixed momentum/position space representation we have
\begin{equation}
    \begin{aligned}
    \hat{H}=\sum_{{\bm k}} \sum_{m,\nu} E^\nu_{m,{\bm k}} c_{m{\bm k}}^{\nu\dagger} c^\nu_{m{\bm k}}+\sum_{{\bm R}}\frac{1}{2} U_0 \hat{N}_{\bm R}(\hat{N}_{\bm R}-8),
    \end{aligned}
    \label{eq:HTBGHubbard}
\end{equation}
where $c_{m{\bm k}}^{\nu\dagger}$ creates an electron in a Bloch spinor eigenstate with band $m$ and flavor (spin and valley) $\nu$; in the position space representation, $c_{n{\bm R}}^{\nu\dagger}$ creates an electron of orbital $n$ and flavor $\nu$ at site ${\bm R}$; $n^\nu_{n{\bm R}}=c_{n{\bm R}}^{\nu\dagger}c_{n{\bm R}}^{\nu}$ is the density operator, and $\hat{N}_{\bm R}= \sum_{n,\nu}n^\nu_{n{\bm R}}$ is the local density at site $R$.

Quantum embedding methods such as dynamical mean field theory~\cite{Georges96,Chen22,Beck_2022} can utilize mixed momentum-position representations.  %, where important interaction correlations are relatively local. 
In these methods an approximation to the electron self energy is obtained from a quantum impurity model defined from a local Green function $G^{{\bm R}}(\omega)$ obtained by downfolding of the k-space Green function $G({\bm k}, \omega)$ using  a projection operator $P$,
\begin{equation}
    G^{{\bm R}}_{n n^{\prime}}\left( \omega\right)=\frac{1}{N}\sum_{{\bm k},mm^{\prime}} (P^{\dagger})_{nm}^{{\bm R}{\bm k}} G_{m m^{\prime}}\left({\bm k},\omega \right) P_{m^{\prime}n^{\prime}}^{{\bm R}{\bm k}}.
    \label{eq:downfold}
\end{equation}

In the case of current interest the projection operator is precisely the unitary transformation used to define the Wannier function, $P_{m n}^{{\bm R}{\bm k}}= e^{-i{\bm k}\cdot{\bm R}} M_{m n}({\bm k})$, and the k-space Green function in the band basis is defined as:
\begin{equation}
    G({\bm k}, \omega)=\left(\omega+\mu-\hat{H}_0({\bm k})-\hat{\Sigma} ({\bm k}, \omega)\right)^{-1},
    \label{eq:k-greenfunction}
\end{equation}
where $\hat{H}_0({\bm k})$ is a $8 \times 8$ diagonal matrix representing the noninteracting Hamiltonian and $\mu$ is the chemical potential. The band-basis self energy, $\Sigma ({\bm k}, \omega)$, is given by upfolding the local self energy via 
\begin{equation}
    \Sigma_{m m^{\prime}}\left({\bm k}, \omega\right)=\frac{1}{N}\sum_{n n^{\prime}{\bm R}} P_{m n}^{{\bm R}{\bm k}} \Sigma^{{\bm R}\mathbf{R^\prime}}_{n n^{\prime}}( \omega)(P^{\dagger})_{n^{\prime} m^{\prime}}^{\mathbf{R^\prime}{\bm k}}.
    \label{eq:upfold}
\end{equation}

The dynamical mean field approximation involves collapsing the full position-dependent Self energy to a site local function: $\Sigma^{{\bm R},\mathbf{R^\prime}}\rightarrow \Sigma^{{\bm R}}\delta_{{\bm R},\mathbf{R^\prime}}$ in the wannier basis. Note however the $\Sigma_{m m^{\prime}}({\bm k}, \omega)$ retains momentum-dependence in the band basis. $\Sigma^{{\bm R}}$   is computed from an impurity model, specified by the local Green's function defined in Eq.~(\ref{eq:downfold}) and the site-local interaction. The impurity model is a quantum many-body problem (a $1$-space $+$ $1$-time dimensional quantum field theory) which in general requires a numerical solution. Here our aim is to demonstrate the method, so we solve the impurity model in a Hartree-Fock approximation and investigate different broken symmetries. 

As a concrete example we present a comparison of valley polarized and valley coherent states at charge neutrality. The self energy corresponding to a  valley polarized state is written as:
\begin{equation}
    \Sigma_{pol}^{{\bm R}}(\omega)=\Sigma_0 I - \Sigma_1 v_{z}
    \label{eq:self_energy}
\end{equation}
where $I$ is the identity matrix and $v_z$ is the Pauli matrix for the valley degree of freedom. $\Sigma_0$ and $\Sigma_1$ are parameters (frequency-independent within the Hartree-Fock approximation) with $\Sigma^1$ quantifying the broken symmetry. The valley coherent state is a rotation from the valley polarized state $\Sigma^{{\bm R}}_{coh}(\omega)=O \Sigma^{{\bm R}}_{pol}(\omega)O^\dagger$ specified by a rotation matrix $O$. Inspired by previous work \cite{Wannier_obstructions2}, we choose two possible valley coherent states, rotated from the valley polarized state with rotation $O_1=e^{-i \frac{\pi}{4} o_{y} v_{x} }$ and $O_2=e^{-i \frac{\pi}{4} o_{x} v_{x} }$. Here $o, v$ are the Pauli matrix for real space orbital and valley, respectively. We take the transformation to be the same for the two spins.

\begin{figure}[htbp]
	\centering
	\includegraphics[width=1.0\linewidth]{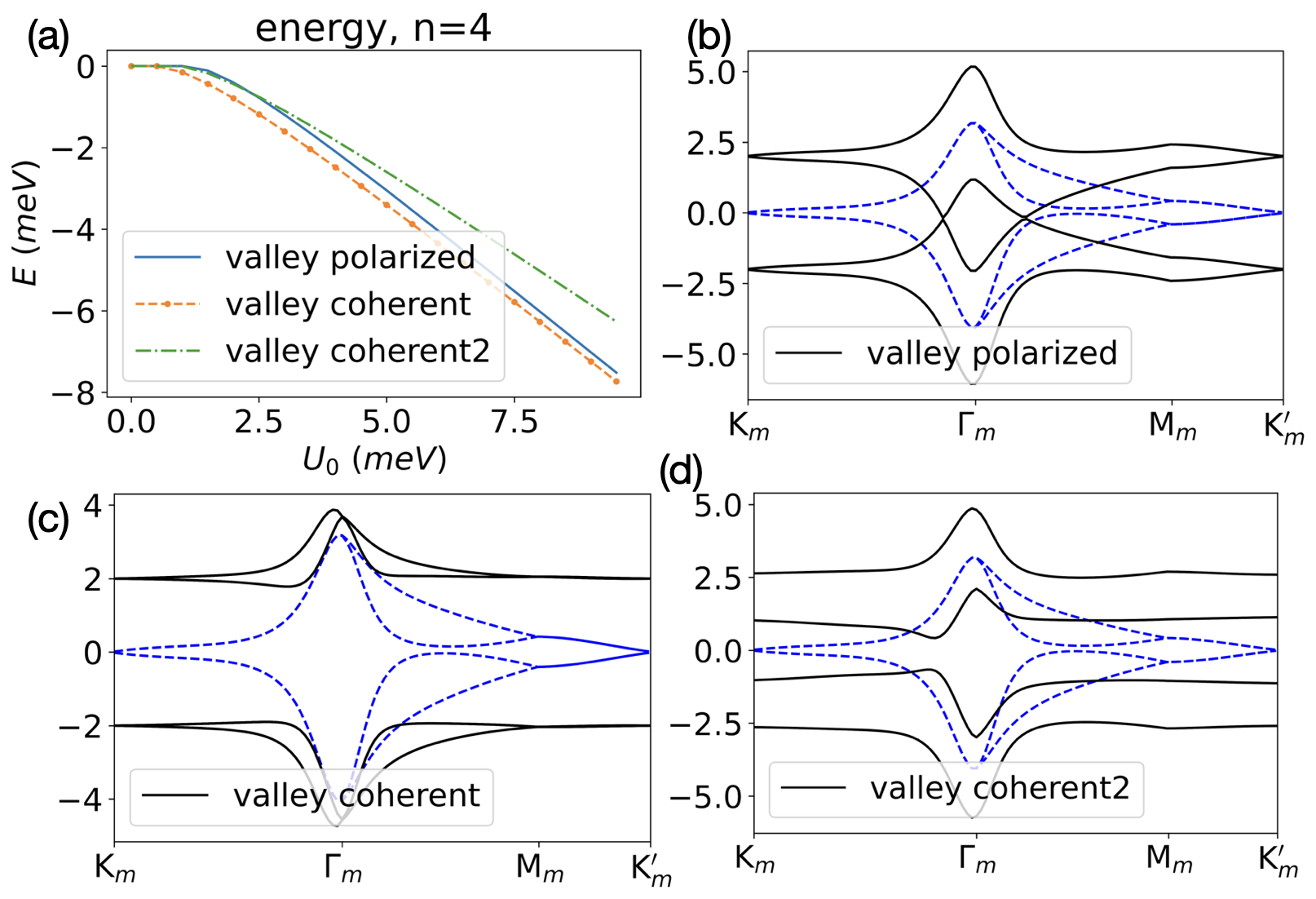}
	\caption{(a) Energy of the ordered state as a function of interaction strength at $n=4$. $E=0$ meV corresponds to the non-interacting state. (b-d) Band structures of different ordered states at $mU_0=2$ meV for $n=4$ (charge neutral filling). The blue dashed line shows the non-interacting band structure.}
	\label{fig:ground_state_n=4}
\end{figure}

The dynamical mean field equations are solved iteratively from an initial trial self energy. We obtained the results for the ground state at charge neutral filling, as shown in Fig.~\ref{fig:ground_state}. The results for other integer fillings are presented in the Appendix. At $n=4$, the valley coherent state I, obtained from the valley polarized state by rotation $O_1$, has a lower energy than the other states; the  energy different is largest at intermediate  interaction. For large interaction, the energy difference between the valley coherent I and the valley polarized state is quite small, being set by a super-exchange process determined by the inverse of the gap size, which is similar for the two states. It is also interesting to note that the valley polarized state is a semimetal at small interaction. If this state could be stabilized it might be susceptible to a further gap opening leading to a  quantum anomalous Hall state. 

\emph{Summary and conclusion---}
This paper has presented a concrete example of the construction and use of a Wannier basis to study correlations in a topologically obstructed system. We show that basis functions reflecting the physical charge density, transforming simply under the relevant point operations, and producing theoretically tractable forms of the interaction Hamiltonian can be constructed. We have further shown that the unavoidable consequences of the topological obstruction (power-law decay of wave functions and non-uniform convergence of momentum space single particle energies with number of Wannier functions) are easily manageable. In particular the amplitude of the power-law tail can be very low so that the charge density is in practice sufficiently localized, while the non-uniform convergence and related issues with representation of symmetry operations in momentum space can be avoided by use of a mixed momentum/ position space representation that occurs naturally in the widely used cluster embedding methods. 

Moreover, we have constructed one representation of a many-body Hamiltonian, which naturally encodes important aspects of the electronic physics, including the physical charge distribution in a unit cell, and provides a representation in which the dominant interaction terms are spatially local. This representation is thus suitable for the investigation of local physics including Mott transitions, orbital ordering, melting transitions and thermal crossover states and certain forms of magnetic ordering. For example, within this local picture, it is easy to construct certain spatially ordered states, such as 120 degree antiferromagnetic order, which is a natural order on the triangular lattice. However, other ordered states (for example the two-sublattice antiferromagnetic state of the honeycomb lattice), do not have a simple and natural expression in this basis. Understanding how to represent the honeycomb-type states in the triangular basis is an interesting open problem. More generally, since quantum many-body Hamiltonians cannot in general be solved exactly, the question of identifying a representation which most compactly encodes the relevant interaction and single-particle physics is an important open problem. 

Last but not least, this work paves the way for the future investigation of correlated states on twisted bilayer graphene~\cite{PhysRevX.10.031034,Wannier_obstructions2,tbg3,tbg5,fang22,Kekul21}, including studies of non-integer fillings~\cite{yonglongxie_fci_21} and temperature dependence~\cite{Pomeranchuk1,Pomeranchuk2} and also sets the stage for investigations of strong correlation effects in other systems with Chern bands and topological obstructions.

\begin{acknowledgments}
We thank Oskar Vafek for discussions on Wannier functions of topological bands and Nan Cheng for discussions on symmetry. J.C., J.Z. and A.J.M acknowledge support from the NSF MRSEC program through the Center for Precision-Assembled Quantum Materials (PAQM) - DMR-2011738. The Flatiron Institute is a division of the Simons Foundation.
\end{acknowledgments}

\appendix
\section{Illustration of Brillouin zone folding}
\begin{figure}[htbp]
	\centering
	\includegraphics[width=0.7\linewidth]{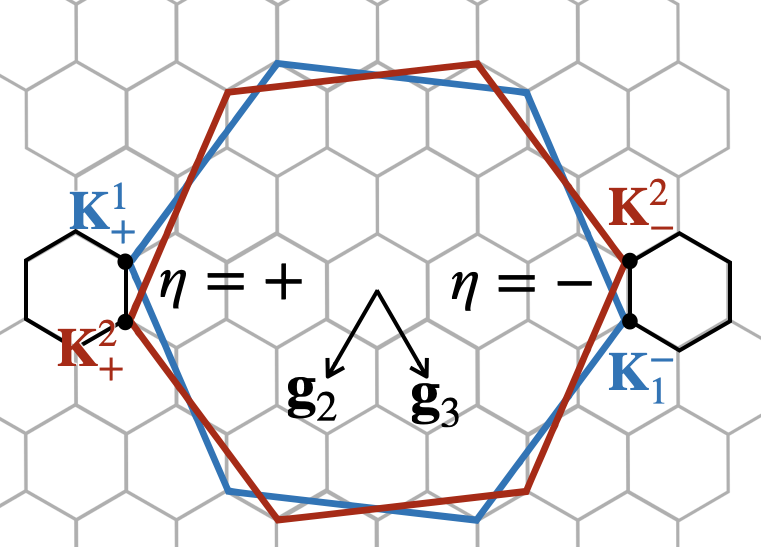}
	\caption{Illustration of Brillouin zone folding in TBG. The large blue and red hexagons represent the first Brillouin zones of graphene layers 1 and 2, and the small hexagon is the moir\'e Brillouin zone of TBG.}
	\label{brillouin_zone}
\end{figure}

Here we show our convention of the Brillouin zone folding.  ${\bm k}^1_{\eta}$ and ${\bm k}^2_{\eta}$ are Dirac points of layer 1 and 2 at valley $\eta$. ${\bm k}^1_{\eta}=R_{-\frac{\theta}{2}}{\bm k}_{\eta},{\bm k}^2_{\eta}=R_{\frac{\theta}{2}}{\bm k}_{\eta}$, where ${\bm k}_{\eta}$ is the un-rotated Dirac point. $k_D=|{\bm k}^1_{\eta}-{\bm k}^2_{\eta}|$. $\mathbf{g}_{1}=0$, $\mathbf{g}_{2}=\sqrt{3}k_D (-\frac{1}{2},-\frac{\sqrt{3}}{2})$, and $\mathbf{g}_{3}=\sqrt{3}k_D (\frac{1}{2},-\frac{\sqrt{3}}{2})$.
\section{Symmetry of Bloch States}
Here we discuss the symmetry $C_{2} \mathcal{T}$, $C_3$, and $C$ for bloch states.

$C_{2} \mathcal{T}=\sigma_{x} \hat{R}_{z}(\pi)\mathcal{K}$, where $\mathcal{K}$ is the complex conjugation operator,  $\hat{R}_{z}(\varphi)=\exp \left[-\varphi {\bm r} \times \partial_{{\bm r}}\right]$, and the Pauli matrix $\sigma$ acts on the sublattice A,B index. We have
\begin{equation}
    \begin{aligned}
        &(C_{2} \mathcal{T})^{-1}H^{\eta}({\bm r})C_{2} \mathcal{T}=H^{\eta}({\bm r})\\
        &\Rightarrow \sigma_{x} \hat{H}^{\eta *}(-{\bm r})  \sigma_{x} =\hat{H}^{\eta}({\bm r}),\\
        &C_2T\psi_{m{\bm k}}^{\eta}({\bm r})=\sigma_{x} \psi_{m{\bm k}}^{\eta *}(-{\bm r})=e^{i \alpha^\eta_{m{\bm k}}} \psi_{m{\bm k}}^{\eta}({\bm r})
        \label{eq:C2T}
    \end{aligned}
\end{equation}
The last equation means that if $\psi_{m{\bm k}}^{\eta}({\bm r})$ is an eigenstate with energy $E_{{\bm k}}$, $C_2T\psi_{m{\bm k}}^{\eta}({\bm r})= \sigma_{x} \psi_{m{\bm k}}^{\eta *}(-{\bm r})$ will also be an eigenstate with the same energy at the same ${\bm k}$ point. For general ${\bm k}$ points where two bands are not degenerate,  $C_2T\psi_{m{\bm k}}^{\eta}({\bm r})$ and $\psi_{m{\bm k}}^{\eta}({\bm r})$ should be the same up to a phase factor $e^{i \alpha^\eta_{m{\bm k}}}$, which depends on gauge choice.

The $C_3$ operator is defined as $C_{3}=W^\dagger({\bm r}) P \hat{R}_{z}(2 \pi / 3)W({\bm r})$, with $W({\bm r})=\operatorname{diag}\left[e^{-i {\bm k}_{\eta}^{1} \cdot {\bm r}},e^{-i {\bm k}_{\eta}^{1} \cdot {\bm r}}, e^{-i {\bm k}_{\eta}^{2} \cdot {\bm r}}, e^{-i {\bm k}_{\eta}^{2} \cdot {\bm r}}\right]$, $P=\text{diag}[e^{i\eta 2\pi/3}, e^{-i\eta  \frac{2\pi}{3}},e^{i\eta \frac{2\pi}{3}},e^{-i\eta  \frac{2\pi}{3}}]=e^{i\frac{2\pi}{3}\sigma_z v_z}$. We have
\begin{equation}
	\begin{aligned}
	   &(C_{3} )^{-1}H^{\eta}({\bm r})C_{3} =H^{\eta}({\bm r})\\
        &\Rightarrow \hat{R}_{z}^\dagger P^\dagger \left (W({\bm r})H^{\eta}({\bm r})W^\dagger({\bm r})\right)P\hat{R}_{z}=W({\bm r})H^{\eta}({\bm r})W^\dagger({\bm r}) \\
        \end{aligned}
\end{equation}
The $C_3$ symmetry defined here has an additional matrix $W({\bm r})$ compared to previous papers \cite{chiral,Vafek_2021}. This is because the ${\bm k}$ point in the Hamiltonian in Eq.~(\ref{eq:Hamiltonian}) is measured from the $\mathbf{\Gamma}$ point instead of ${\bm k}^1_{\eta}$ or ${\bm k}^2_{\eta}$. $W({\bm r})H^{\eta}({\bm r})W^\dagger({\bm r})$ can be thought as an unitary transformation from $H^{\eta}({\bm r})$. Correspondingly, the eigenstate behaves as
\begin{equation}
	\begin{aligned}
        &C_3\psi_{m{\bm k}}^{\eta}({\bm r})=W^\dagger({\bm r}) P W(C_3^{-1}{\bm r})  \psi_{m{\bm k}}^{\eta }(C_3^{-1}{\bm r})=e^{i\beta^\eta_{mk}}\psi^{\eta}_{mC_3 {\bm k}}(r)\\
		&\Rightarrow PW(C_3^{-1}{\bm r})  \psi_{m{\bm k}}^{\eta }(C_3^{-1}{\bm r})=e^{i\beta^\eta_{m{\bm k}}}W({\bm r})\psi^{\eta}_{mC_3{\bm k}}({\bm r})
	\end{aligned}
	\label{eq:C3}
\end{equation}

At the chiral limit, the chiral particle-hole symmetry is defined as $C=\sigma_{z}$. We have
\begin{equation}
    \begin{aligned}
    &(C)^{-1}H^{\eta}({\bm r})C =-H^{\eta}({\bm r})\\
    &\Rightarrow \sigma_z H^{\eta}({\bm r})\sigma_z=-H^{\eta}({\bm r})\\
    &C\psi_{m {\bm k}}^{\eta}({\bm r})=\sigma_z \psi_{m {\bm k}}^{\eta}({\bm r})=e^{i 
    \zeta^\eta_{{\bm k}}} \psi_{\bar{m},{\bm k}}^{\eta}({\bm r}),
    \end{aligned}
\end{equation}
where $\bar{m}$ refers to the band not labelled by $m$.

\section{Wannier Function Construction\label{sec:AppendixC}}
For each spin and valley, two Wannier orbitals $n=1,2$ at site ${\bm R}$ are constructed from the Fourier transformation of two Bloch states $m=1,2$:
\begin{equation}
	\begin{aligned}
	&w^\eta_{n{\bm R}}({\bm r})=\frac{1}{\sqrt{N}} \sum_{{\bm k}} e^{-i {\bm k} \cdot {\bm R}}\sum_{m=1,2} \psi^\eta_{m {\bm k}}({\bm r}) M^\eta_{m n}({\bm k})\\
	&M^\eta ({\bm k})=\left(\begin{array}{cc}a^\eta_{k} & e^{i \phi^\eta_{k}} b_{k}^{\eta*} \\ b^\eta_{k} & -e^{i \phi^\eta_{k}} a_{k}^{\eta *}\end{array}\right),~|a_{k}^{\eta }|^{2}+|b^\eta_{k}|^{2}=1
	\label{eq_app:wannier}
\end{aligned}
\end{equation}
where ${\bm R}=m_1\mathbf{a}_{M1}+m_2\mathbf{a}_{M2}$ ($m_1,m_2\in Z$) is the lattice site, and $\mathbf{a}_{M1}, \mathbf{a}_{M1}$ are the moir\'e lattice vectors, as shown in Fig.~\ref{fig:real_space}(a). $M^\eta({\bm k})$ is a $2 \times 2$ unitary matrix that guarantees the orthogonality of two Wannier functions.

\begin{figure}[htbp]
	\centering
	\includegraphics[width=1\linewidth]{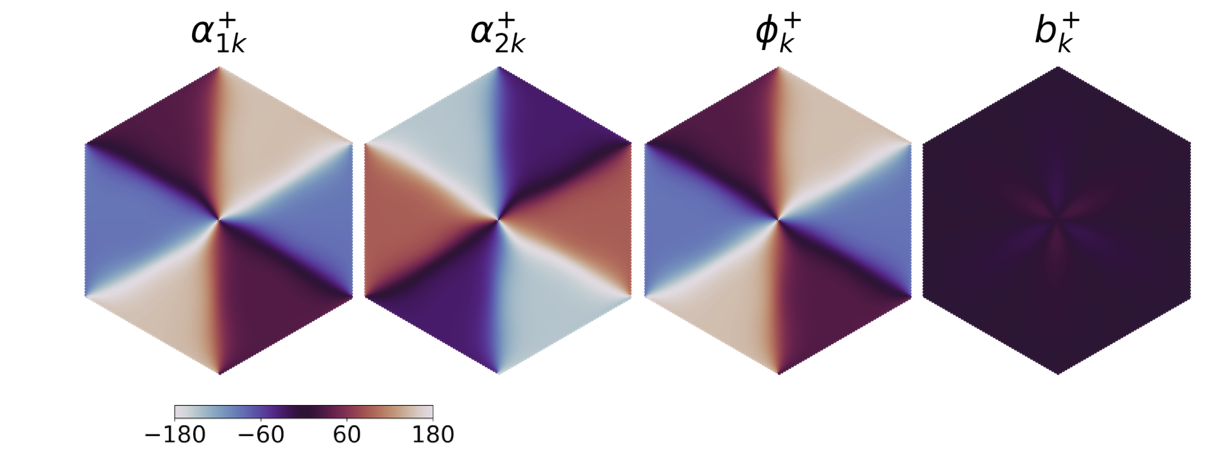}
	\caption{$\alpha_{1 {\bm k}}^{+}$, $\alpha_{2 {\bm k}}^{+}$, $\phi_{ {\bm k}}^{+}$, and the phase of $b_{k}^{+}$ when it is not at chiral limit.} 
	\label{fig:phase}
\end{figure}

Each Bloch state is a spinor of four components (A1, B1, A2, B2), and we first fix the phase of Bloch states in eq.~\ref{eq_app:wannier} by requiring that their B1 components, $\psi^{\eta B1}_{1{\bm k}} ({\bm r})$ and $\psi^{\eta B1}_{2{\bm k}} ({\bm r})$, are real and positive for each ${\bm k}$ point at ${\bm r}=0$. This also fixes the symmetry phase $\alpha_{m {\bm k}}^{\eta},\beta_{m{\bm k}}^{\eta}$ and $\zeta_{{\bm k}}^{\eta}$ defined in eq.~\ref{eq:bloch_symmetry} and  \ref{eq:C}. $e^{i \beta_{m {\bm k}}^{\eta}}=e^{-i 2 \pi / 3\eta}$, $e^{i \zeta_{\mathrm{k}}^{n}}=-1$, and $\alpha_{m {\bm k}}^{\eta}$ is shown in Fig.\ref{fig:phase}.

To require Wannier functions have the following symmetries,
\begin{equation}
	\begin{aligned}
	    C_3 w_{n{\bm R}}^{\eta}({\bm r}) &=e^{-i\frac{2\pi}{3}n\eta}w_{nC_3{\bm R}}^{\eta}({\bm r})\\
	   C_{2} \mathcal{T} w^\eta_{1{\bm R}}({\bm r})&=\sigma_{x}w^{\eta *}_{1{\bm R}}(-{\bm r})=w^\eta_{2,-{\bm R}}({\bm r})
	\end{aligned}
\end{equation}
$a_{k}^{\eta}$, $b_{k}^{\eta}$ and $e^{i \phi_{k}^{\eta}}$ should satisfy:
\begin{equation}
	\begin{aligned}
	   C_3:~& a_{k}^{\eta}=a_{C_3k}^{\eta},b_{k}^{\eta}=b_{C_3k}^{\eta},\phi_{k}^{\eta}=e^{-i2\pi/3}\phi_{C_3k}^{\eta}\\
	   C_{2} \mathcal{T}:~&b_{k}^{\eta}=-e^{i \alpha_{2k}^{\eta }} e^{-i \phi^\eta_{k}} a^\eta_{k}\\
	   &\phi^\eta_{k}=\frac{\alpha_{1k}^{\eta }+\alpha_{2k}^{\eta }\pm\pi}{2}
	\end{aligned}
	\label{eq:phase_symmetry}
\end{equation}
 
\begin{figure}[htbp]
	\centering
	\includegraphics[width=1.0\linewidth]{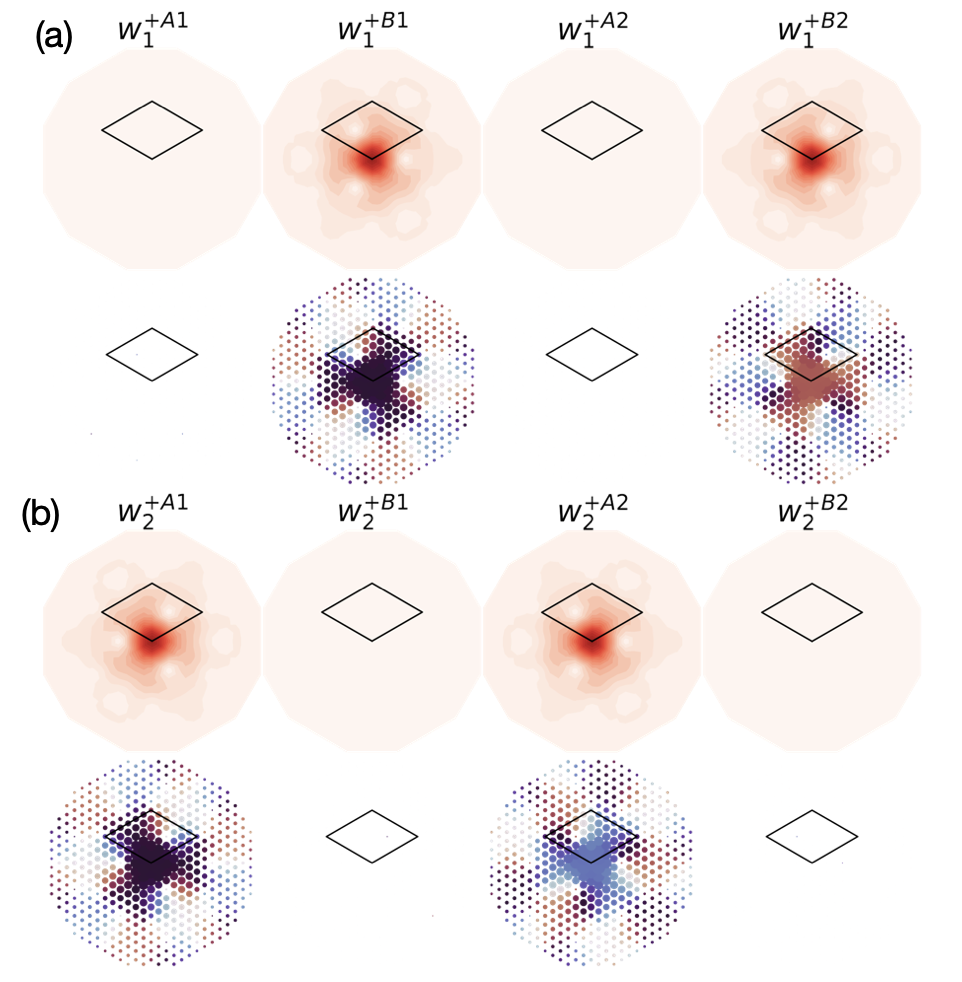}
	\caption{Two Wannier functions $w_{1 {\bm R}}^{+}({\bm r})$ and $w_{2 {\bm R}}^{+}({\bm r})$ for valley $+$ at ${\bm R}=0$ at chiral limit ($u_0=0$). $w_{1 {\bm R}}^{+A1}({\bm r})=w_{1 {\bm R}}^{+A2}({\bm r})=w_{2 {\bm R}}^{+B1}({\bm r})=w_{2 {\bm R}}^{+B2}({\bm r})=0$.}
	\label{fig:wannier_tri_chiral}
\end{figure}

We take $a_k^{\eta}=\frac{1}{\sqrt{2}}$, and treat the branch cut in $\phi_{k}$ carefully such that $\phi_{k}^{\eta}=e^{-i2\pi/3}\phi_{C_3k}^{\eta}$. $\phi_k$ and the phase of $b_{k}^{\eta}$ are shown in Fig.~\ref{fig:phase}. At the chiral limit, $\alpha_{1k}^{\eta }=\alpha_{2k}^{\eta }\pm\pi$, and Eq.~(\ref{eq:phase_symmetry}) implies $a_k^{\eta}=\frac{1}{\sqrt{2}}$ and $\phi_{k}^{\eta}=\alpha_{1k}^{\eta }$. The Wannier functions for the chiral limit case are shown in Fig.~\ref{fig:wannier_tri_chiral}.

\section{Ordered state}
In the main text we show the results for charge neutral fillings ($n=4$). At other integer fillings, the calculation is performed in a similar way.

\begin{figure}[htbp]
	\centering
	\includegraphics[width=1.0\linewidth]{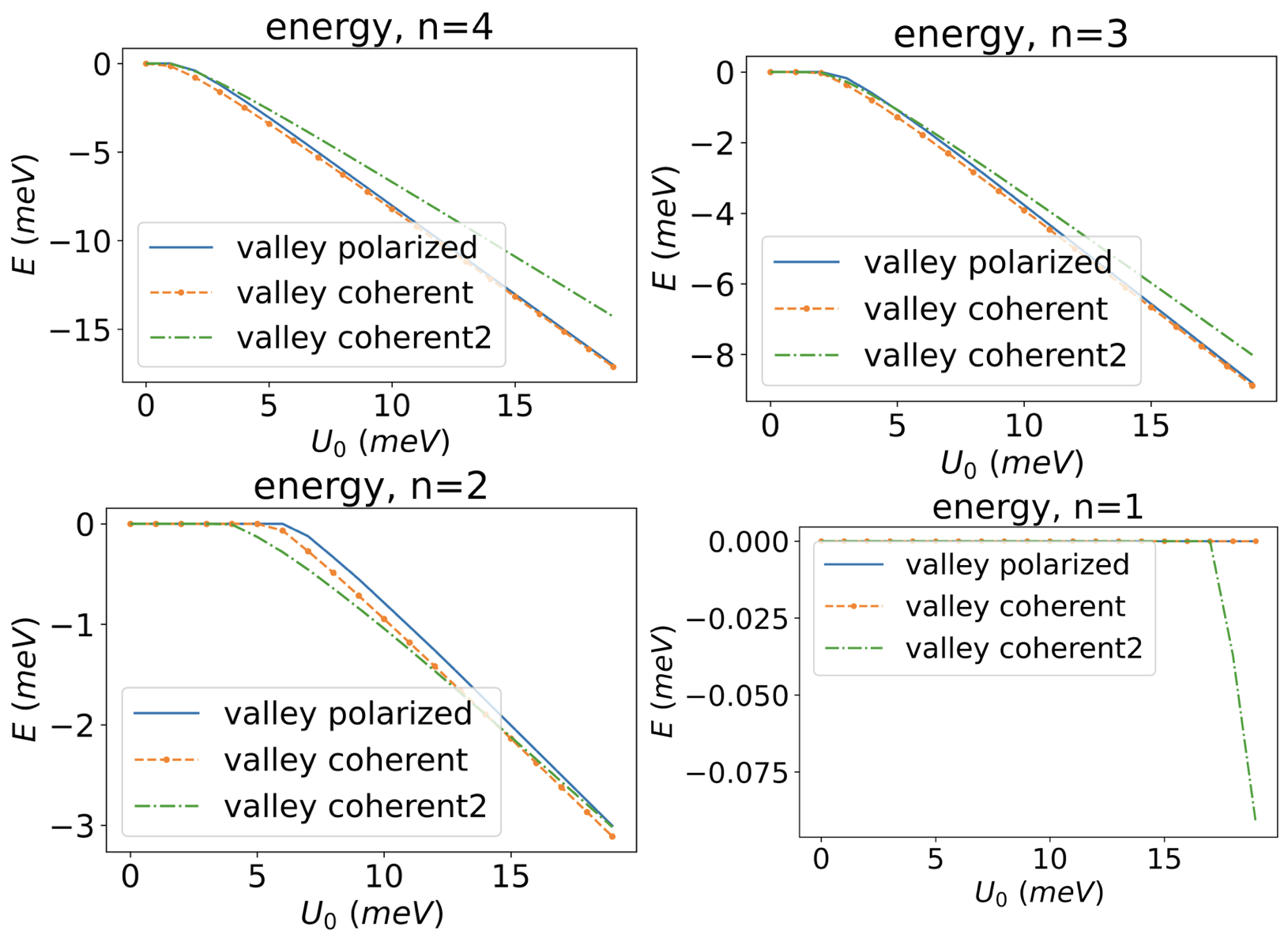}
	\caption{Energy of the ordered state as a function of interaction strength at integer fillings. $E=0$ meV corresponds to non-interacting state. }
	\label{fig:ground_state}
\end{figure}

For valley polarized state, at $n=3$, we assume $\langle n^+_{1{\bm R}\uparrow}\rangle=\langle n^+_{2{\bm R} \uparrow}\rangle=\langle n^+_{1{\bm R} \downarrow}\rangle=n / 8+m$, and $\langle n^+_{2{\bm R}\downarrow}\rangle=\langle n^-_{1{\bm R} \uparrow}\rangle=\langle n^-_{2{\bm R}\uparrow}\rangle=\langle n^-_{1{\bm R}\downarrow}\rangle=\langle n^-_{2{\bm R}\downarrow}\rangle=n / 8-m$; at $n=2$, we assume $\langle n^+_{1{\bm R}\uparrow}\rangle=\langle n^+_{2{\bm R} \uparrow}\rangle=n / 8+m$, and $\langle n^+_{1{\bm R} \downarrow}\rangle=\langle n^+_{2{\bm R}\downarrow}\rangle=\langle n^-_{1{\bm R} \uparrow}\rangle=\langle n^-_{2{\bm R}\uparrow}\rangle=\langle n^-_{1{\bm R}\downarrow}\rangle=\langle n^-_{2{\bm R}\downarrow}\rangle=n / 8-m$; at $n=1$ we assume $\langle n^+_{1{\bm R}\uparrow}\rangle=n / 8+m$, and $\langle n^+_{2{\bm R} \uparrow}\rangle=\langle n^+_{1{\bm R} \downarrow}\rangle=\langle n^+_{2{\bm R}\downarrow}\rangle=\langle n^-_{1{\bm R} \uparrow}\rangle=\langle n^-_{2{\bm R}\uparrow}\rangle=\langle n^-_{1{\bm R}\downarrow}\rangle=\langle n^-_{2{\bm R}\downarrow}\rangle=n / 8-m$. Here $m$ is the magnetization.

The valley coherent state is a rotation from the valley polarized state, $a'^\dagger=a^\dagger O$. We choose two possible valley coherent states, rotated from valley polarized state with rotation $O_1=e^{-i \frac{\pi}{4} o_{y} v_{x}}$ and $O_2=e^{-i \frac{\pi}{4} o_{x} v_{x}}$ where $o, v$ are the Pauli matrix for real space orbital and valley, respectively. The results are shown in Fig.~\ref{fig:ground_state}

\nocite{*}

\bibliography{apssamp}% Produces the bibliography via BibTeX.

\end{document}